\documentclass[aps,twocolumn,floats,superscriptaddress,prd,nofootinbib]{revtex4}
\usepackage{graphicx}
\usepackage{bm}
\usepackage{amssymb}
\usepackage{amsmath}
\usepackage{cancel}
\usepackage{hyperref}
\usepackage{amsmath}

\usepackage{epstopdf}
\usepackage{natbib}
\usepackage{epsfig}

\newcommand{\avg}[1]{\ensuremath{\langle #1 \rangle}}
\newcommand{\bma}{\begin{math}}
\newcommand{\ema}{\end{math}}
\newcommand{\beq}{\begin{equation}}
\newcommand{\eeq}{\end{equation}}
\newcommand{\beqa}{\begin{eqnarray}}
\newcommand{\eeqa}{\end{eqnarray}}
\newcommand{\bc}{\begin{center}}
\newcommand{\ec}{\end{center}} 
\newcommand{\bit}{\begin{itemize}}
\newcommand{\eit}{\end{itemize}}

\font\BFd=cmmib10
\font\BFt=cmmib10
\font\BFs=cmmib10 scaled 700
\font\BFss=cmmib10 scaled 500

\def\bbox#1{%
\relax\ifmmode
\mathchoice
{{\hbox{\BFd #1}}}
{{\hbox{\BFt #1}}}
{{\hbox{\BFs #1}}}
{{\hbox{\BFss #1}}}
\else \mbox{#1} \fi }

\def\k{{\bbox{k}}}

\def\x{{\bbox{x}}}

\usepackage{color}

\definecolor{darkgreen}{cmyk}{0.85,0.2,1.00,0.2}

\begin{document}

\title{Primordial Non-Gaussianity and Reionization}

\author{Adam Lidz}
\affiliation{Department of Physics \& Astronomy, University of Pennsylvania, 209 South 33rd Street, Philadelphia, Pennsylvania 19104}

\author{Eric J Baxter}
\affiliation{Department of Astronomy \& Astrophysics, University of Chicago, Chicago Illinois 60637}

\author{Peter Adshead}
\affiliation{Kavli Institute for Cosmological Physics,  Enrico Fermi Institute, University of Chicago, Chicago, Illinois 60637}

\author{Scott Dodelson}
\affiliation{Fermilab Center for Particle Astrophysics, Fermi National Accelerator Laboratory, Batavia, Illinois 60510-0500}
\affiliation{Kavli Institute for Cosmological Physics,  Enrico Fermi Institute, University of Chicago, Chicago, Illinois 60637}
\affiliation{Department of Astronomy \& Astrophysics, University of Chicago, Chicago Illinois 60637}

\begin{abstract}
The statistical properties of the primordial perturbations contain
clues about the origins of those fluctuations.  Although the Planck collaboration has recently obtained tight constraints on primordial non-gaussianity from
cosmic microwave background measurements,
it is still worthwhile to mine upcoming data sets in effort to place independent or competitive limits.
The ionized bubbles that formed at redshift
$z\sim6-20$ during the Epoch of Reionization are seeded by primordial overdensities, and so the statistics
of the ionization field at high redshift are related to the statistics
of the primordial field. Here we model the effect of primordial
non-gaussianity on the reionization field. The epoch and duration of
reionization are affected as are the sizes of the ionized bubbles, but
these changes are degenerate with variations in the properties of the ionizing sources and 
the surrounding intergalactic medium. A more promising signature is the power spectrum of the spatial
fluctuations in the ionization field, which may be probed by upcoming 21 cm surveys. This has
the expected $1/k^2$ dependence on large scales, characteristic of a biased tracer of the matter field.
We project how well upcoming 21 cm observations will be able to disentangle this signal from
foreground contamination. Although foreground cleaning inevitably removes the 
large-scale
modes most impacted by primordial non-gaussianity, we find
that primordial non-gaussianity
can be separated from foreground contamination for a narrow range of length
scales. In principle, futuristic redshifted 21 cm surveys may allow constraints competitive with Planck.
\end{abstract}

\maketitle

\section{Introduction}

The standard cosmological model makes predictions that have been
confirmed in a variety of arenas, from the cosmic microwave background
to large surveys of galaxies. The model contains three mysteries,
elements that are not part of the Standard Model of particle physics:
dark matter, dark energy, and inflation. Inflation currently plays the
role of providing the seeds of structure, and the simplest
inflationary models predict that the perturbations responsible for
this structure were drawn from a gaussian distribution. Evidence for
primordial non-gaussianity then would speak to either a more complex
inflation model or an alternative in which early acceleration does not
occur. Either would be fascinating and probe physics operating at the 
earliest moments in the history of our Universe.

The Planck Collaboration~\cite{Ade:2013ydc} has, however, recently -- as we were completing this work -- 
placed stringent constraints
on primordial non-gaussianity, by determining a robust upper limit to the 3-point
function of the cosmic microwave background~\cite{Komatsu:2002db,Creminelli:2005hu,Smith:2009jr}.
These results provide support for the simplest models of inflation with gaussian primordial fluctuations; nature may not provide us with this potential handle on the physics of inflation and
alternatives. Given the enormous significance of a detection of primordial non-gaussianity, it is nonetheless
worth pursuing additional observational constraints. Since the cosmic microwave background
is relatively unscathed by gravitational, non-linear effects, it will be challenging to improve
on the Planck constraint or to confirm it independently. The two-point function of biased
tracers (galaxies, clusters, etc.)~\cite{Dalal:2007cu} may, however, be able to provide competitive
constraints.

While galaxy surveys have been the premier method of studying the large scale
structure of the Universe, upcoming 21 cm surveys will map the distribution of neutral
hydrogen and therefore provide another handle. Ultimately the large number of Fourier
modes potentially accessible to 21 cm surveys, may allow even more stringent
constraints than possible with galaxy surveys and the cosmic microwave background~\cite{Loeb:2003ya}. In principle, an extremely futuristic
redshifted 21 cm survey could even detect the level of non-gaussianity expected from the simplest {\em single
field models of slow-roll inflation}~\cite{Cooray:2006km}.
The first aim of redshifted 21 cm surveys is, however, to map out the details of the reionization process
at $z\sim6-20$. Motivated in part by the ultimate promise of the redshifted 21 cm line as a probe of primordial 
non-gaussianity, we examine here whether measurements during the reionization epoch might themselves provide
a useful probe. During reionization, initial halo
formation triggered early star formation that produced radiation
sufficient to ionize large bubbles. An important realization from the
past decade of theoretical work is that regions with large scale
over-densities are ionized before typical regions (e.g., \cite{Barkana:2003qk, Furlanetto:2004nh, Iliev:2005sz, Zahn:2006sg}). This results because
small scale halo formation is biased: halos preferentially form in large
scale over-densities.
Given that primordial non-gaussianity affects this
biasing~\cite{Dalal:2007cu}, it  is interesting to study the impact of
non-gaussianity on reionization.

There have been several related studies in the past \cite{Crociani:2008dt,Joudaki:2011sv,Tashiro:2012wr,
Chongchitnan:2013oxa}. Our work has most overlap with Ref.~\cite{Joudaki:2011sv}. These authors focused
on the scale-dependent clustering of the ionized regions, while we additionally quantify the impact
of primordial non-gaussianity on the timing of reionization and the size distribution of ionized
regions, and compare with analytic predictions for the scale-dependent biasing signature. Furthermore, we
consider the impact of foreground cleaning in more detail than previous authors. This may ultimately provide
the largest obstacle for obtaining precise constraints on primordial non-gaussianity from redshifted 21
cm observations, and so we estimate its impact carefully here.

The outline of this paper is as follows.
We study non-gaussianity and reionization here by combining the
technologies developed by Furlanetto et al.~\cite{Furlanetto:2004nh}
to model reionization and the path integral formalism of
Ref.~\cite{Maggiore:2009rx} to quantify the impact of
non-gaussianity. In \S\ref{secfzh}, we review the Furlanetto model and
explain how non-gaussianity impacts the collapse fraction and
therefore the over-density threshold above which a large scale region
will be considered ionized. Then, we carry out semi-numerical
simulations in \S\ref{sec:sim} and show results for the bubble size and
ionization history in non-gaussian models. Then, in
\S\ref{sec:scale_b}, we compute the two-point function of the
ionization field in the simulations and demonstrate that the power
spectrum rises on large scales. Finally, we conclude in
\S\ref{sec:observability} by projecting how well upcoming 21 cm
surveys will be able to measure this feature in the power spectrum in
the presence of astrophysical foregrounds that pollute the large scale
spectrum. We conclude in \S\ref{sec:conclusions}.

\section{The FZH model and Non-Gaussianity}\label{secfzh}

In effort to understand the impact of primordial non-gaussianity on
the reionization process, we will extend the analytic reionization model of
Ref.~\cite{Furlanetto:2004nh} (hereafter `FZH') to the case of non-gaussian initial conditions. 
In order for this work to be self-contained we briefly summarize
the FZH model here, but refer the reader to the original paper
for a complete treatment.
The crux of FZH is that galaxies form first in large scale
overdense regions and that these regions hence reionize
before typical parts of the Universe. Extended Press-Schechter (EPS) theory and the excursion
set formalism~\cite{Bond:1990iw} describe
how halo formation -- and by extension galaxy formation -- is enhanced
in large scale overdense regions, and so we can apply EPS to
model reionization. This technique most faithfully captures large
scale variations in the timing of reionization; we anticipate
that primordial non-gaussianity will have its most dramatic impact
on precisely these large scale variations. Furthermore, upcoming redshifted
21 cm surveys will measure only the large scale features of reionization (e.g. \cite{Furlanetto:2006jb,Lidz:2007az}). 
For these reasons, the FZH model is well-suited for our present purposes.

In order for a region to be reionized,
the number of photons emitted by sources in the region 
must {\em at least} exceed the number of hydrogen atoms contained within the region. Since ionized
atoms can recombine, it takes more than one photon per atom to reionize a patch of the Universe: recent simulations suggest that
a few photons per atom should suffice (e.g. \cite{2009MNRAS.394.1812P}). In the simplest variant of FZH that we follow here, one supposes
that each galaxy, with total host halo mass $M_{\rm gal}$, can ionize a mass of hydrogen (accounting for some average
number of recombinations) proportional to its host halo mass,
\beqa
M_{\rm ion} = \zeta M_{\rm gal}.
\label{eq:mion}
\eeqa
Here $\zeta$ is an ionizing efficiency parameter that depends on the fraction of galactic baryons
that are converted into stars, the number of ionizing photons that are produced per baryon converted
into stars, the fraction of ionizing photons that escape galactic host halos and make it into the
surrounding IGM, the average recombination rate in the IGM, and other factors. 
Plausible values for these quantities yield $\zeta \sim 10$, although with substantial uncertainties (e.g. \cite{Furlanetto:2006jb}). 

We further assume that every dark matter halo above some minimum (total, i.e., dark matter plus baryons) mass, 
$M_{\rm min}$, 
hosts a galaxy. Throughout we assume that $M_{\rm min}$
is set by the mass scale at which the virial temperature is $10^4$ K, above 
which gas can dissipate thermal energy by emitting atomic lines, condense into
the center of the halo, and eventually form stars. It provides a plausible lower
limit for the host halo mass of a galaxy. 
At high redshifts where $\Omega_m(z) \approx 1$, the
corresponding atomic cooling mass scale is~\cite{Barkana:2000fd}:
\beqa
M_{\rm min} = 1.2 \times 10^8 M_\odot \left(\frac{9}{1+z}\right)^{3/2} \left(\frac{T_{\rm vir}}{10^4 K}\right)^{3/2} 
\left(\frac{0.27}{\Omega_{\rm m}}\right)^{1/2}. \nonumber \\
\label{eq:mcool}
\eeqa

Given these assumptions about the ionizing sources and recombinations in the IGM, FZH consider
spheres of varying radius around every point in the IGM. Each gas parcel in the IGM is approximated to be
either completely ionized or completely neutral.
According to Equations \ref{eq:mion} and
\ref{eq:mcool}, if a sufficiently large fraction of the matter contained in a given sphere is collapsed into
galaxy-hosting halos, the region should be ionized by the sources within. In particular, suppose
a region has overdensity $\delta_m$ when the linear density field is smoothed with a (real space) spherical top-hat
of co-moving radius $R_m$, enclosing a Lagrangian mass $M_m = \avg{\rho_{\rm m}} 4 \pi R_m^3/3$, with $\avg{\rho_{\rm m}}$ indicating the
co-moving cosmic mean matter density. Let us further denote the variance of the linear density field
on this smoothing scale by $S_m$, and the variance smoothed on mass scale $M_{\rm min}$ by $S_n$. The condition for the region to be ionized is then:
\beqa
\zeta f_{\rm coll}(S_n|\delta_m, S_m) \geq 1.
\label{eq:deltax}
\eeqa
Here the symbol $f_{\rm coll}$ refers to the collapse fraction in the region, i.e., to the fraction of matter in
the region that is in halos
of mass larger than $M_{\rm min}$.
The minimum overdensity for which this condition is satisfied is given the label $\delta_m = \delta_X$. A given point in the
IGM is considered to be part of an ionized region of size $R_m$ when $R_m$ is the {\em largest} smoothing
scale for which this condition is satisfied.\footnote{FZH describe how this approximately accounts
for the possibility that a region is ionized by a {\em neighboring} cluster of sources.}  In the event that
there is no smoothing scale around a given point for which this criterion is satisfied, the point is considered
to be completely neutral.

With these assumptions, FZH treat reionization -- in the language of excursion set theory -- as a `barrier-crossing' problem. In the excursion set formalism
one considers the behavior of the smoothed field about a point, $\delta_m$, as a function of decreasing smoothing scale or equivalently 
increasing $S_m$; each realization of $\delta_m$ with increasing variance, $S_m$, is said to follow a `trajectory'. The
statistical properties of the ionized regions then follow from considering the probability distribution that
trajectories cross the barrier condition of Equation \ref{eq:deltax} at various smoothing scales. 

In order to extend this treatment to the case of non-gaussian initial conditions, two steps are hence involved.
First, we need to calculate the conditional collapse fraction in an overdense region, $f_{\rm coll}(S_n|\delta_m, S_m)$,
for non-gaussian models. This in turn defines the reionization barrier -- $\delta_X(S_m)$ -- through Equation \ref{eq:deltax}. Second, we
need to consider non-gaussian modifications to the probability distribution for trajectories to cross this barrier.
A technical challenge with both of these steps is that non-gaussianity induces distinctive correlations between
$\delta_m$ at different smoothing scales, which must be incorporated into these calculations. This 
complicates things considerably in
comparison to the case considered by FZH. In the usual FZH model, barrier crossing probability distributions are calculated
using a top-hat smoothing filter in $k$-space and assuming gaussian initial conditions.\footnote{Note that in FZH,
and many other applications of excursion set theory, formulas for the collapse fraction and other quantities of interest are {\em derived} assuming a top-hat filter
in $k$-space, yet {\em applied using a top-hat in real space}. More specifically, the collapse fraction formulas
involve the variance of the linear density field smoothed on various scales, and these are generally calculated
using a real space top-hat although the formulas themselves are derived using $k$-space filters.}
With these assumptions, different steps in a trajectory are uncorrelated: the future evolution of a trajectory with increasing
variance is independent of its past history. This is the defining characteristic of a Markov process. Fortunately, Maggiore, Riotto and De Simone \cite{Maggiore:2009rv,Maggiore:2009rw,Maggiore:2009rx,DeSimone:2010mu,DeSimone:2011dn} developed a path integral formulation of excursion set theory which allows one to study departures from the Markov case, including the mode-couplings induced by primordial non-gaussianity (see also the related works \cite{D'Amico:2010ta, Paranjape:2011ak, Musso:2012ch, Musso:2012qk}).

\subsection{Non-gaussian Models}

Throughout the present work, we specify to the special case of local models of primordial non-gaussianity. In these models,
the primordial curvature perturbation (on scales smaller than the horizon) is parametrized by:
\beqa
\Phi(\x) = \phi_G(\x) + f_{\rm NL} \left[\phi_G^2(\x) - \avg{\phi_G^2(\x)}\right].
\label{eq:png_potential}
\eeqa
Here $\phi_G(x)$ is a gaussian random field and $f_{\rm NL}$ characterizes the strength of the non-gaussianity.
This form produces a bispectrum that is peaked for
squeezed triangles -- i.e., for $k$-space triangles in which one wavevector has much smaller magnitude
than the other two. Although the above form of primordial non-gaussianity is only one of many possibilities,
it is the most well-studied, and is expected for a wide range of different scenarios, such as multi-field
inflationary models. Recent Planck results provide tight constraints on these models, finding
$f_{\rm NL}^{\rm local} = 2.7 \pm 5.8$ at $68\%$ confidence level~\cite{Ade:2013ydc}. We will nonetheless consider significantly 
larger values for $f_{\rm NL}$ in order to best illustrate the effects of primordial non-gaussianity on
reionization. In addition to the local-type non-gaussianity considered here, it is also common to consider
equilateral type non-gaussianity, which peaks for triangles with $k_1 \approx k_2 \approx k_3$, as well 
`folded' and `orthognal' triangle configurations (see e.g. \cite{Ade:2013ydc}). It may also be possible
to improve on the Planck collabroation's current $68\%$ confidence constraints on these other types of
non-gaussianity, $f_{\rm NL}^{\rm equil} = 42 \pm 75$, and $f_{\rm NL}^{\rm ortho} = -25 \pm 39$~\cite{Ade:2013ydc}, but we don't
consider this explicitly here (although see \cite{Chongchitnan:2013oxa}).

\subsection{The Reionization Barrier in Non-gaussian Models}

We first consider how the reionization barrier is modified in models with primordial non-gaussianity. Adshead et al.~\cite{Adshead:2012hs}
and D'Aloisio et al.~\cite{DAloisiobias} calculate the conditional collapse fraction in models with primordial non-gaussianity using a 
top-hat filter in k-space and the path integral formulation of the excursion set formalism from \cite{Maggiore:2009rx}.  
For the special case of spherical collapse, Adshead et al.'s Equation (46) \cite{Adshead:2012hs} gives an approximate form for
the collapse fraction in a region
of large-scale overdensity $\delta_m$. Retaining a few terms, dropped in the large scale
limit that was taken in Adshead et al. Equation (46), we have:
\begin{align}
& f_{\rm coll}(S_n|\delta_m, S_m) \approx {\rm erfc}\left[\frac{\delta_c(z) - \delta_m}{\sqrt{2 (S_n - S_m)}}\right] \nonumber \\
& + \Bigg\{ \frac{\avg{\delta_n^3} - \avg{\delta_m^3} + 3 \avg{\delta_m^2 \delta_n} - 3 \avg{\delta_n^2 \delta_m}}{3 \sqrt{2 \pi} (S_n - S_m)^{3/2}}
\left[\frac{(\delta_c(z) - \delta_m)^2}{S_n - S_m}- 1\right] \nonumber \\ 
& + \left(\avg{\delta_n^2 \delta_m} + \avg{\delta_m^3} - 2 \avg{\delta_m^2 \delta_n}\right) \frac{\delta_m}{S_m} \frac{\delta_c(z) - \delta_m}{\sqrt{2 \pi (S_n - S_m)^3}} \Bigg\} \nonumber \\ 
& \times {\rm exp}\left[-\frac{(\delta_c(z) - \delta_m)^2}{2 (S_n - S_m)}\right].
\label{eq:fcoll_png}
\end{align}
Here $\delta_c(z) = 1.686 D(0)/D(z)$ is the critical overdensity for spherical collapse scaled
to $z=0$ and $D(z)$ is the linear growth factor. The quantity $S_n$
is the variance of the linear density field, when 
the density field is smoothed on the
scale $M_{\rm min}$, $S_m$ is the linear variance smoothed on large scale ($M_m$), 
and $\avg{\delta_n^3}$ is the skewness at smoothing scale $M_{\rm min}$. Similarly, $\avg{\delta_m^3}$ is the
skewness smoothed on the large scale $M_m$, while $\avg{\delta_n^2 \delta_m}$ and $\avg{\delta_m^2 \delta_n}$
are cross terms that also arise in non-gaussian models.
Each of these quantities
is linearly evolved to the present day ($z=0$). The first term is the usual
result for gaussian initial conditions from Press-Schechter theory,
while the second and third terms are approximate modifications
for a Universe with non-gaussian initial conditions. As detailed in Ref.~\cite{Adshead:2012hs}, the non-gaussian
corrections -- i.e., the second and third terms above -- are derived assuming $\delta_c^2(z) \gg S_m$ and including only dominant three-point correlator terms.
This expression is equivalent to D'Aloisio et al.'s Equation (32) \cite{DAloisiobias} in the limit $\delta_c^2(z) \gg S_m$, $\delta_c(z) \gg \delta_m$,
and ignoring their last term (i.e., the last line of their Equation (32), 
proportional to their `$\mathcal{C}$'), which is negligible for the case considered here.

For positive $f_{\rm NL}$, the conditional collapse fraction of Equation \ref{eq:fcoll_png} is enhanced compared to the case of gaussian initial conditions. This results because positive $f_{\rm NL}$ enhances the high density tail of the probability distribution function and thereby boosts the
chance of being above the collapse threshold, $\delta_c(z)$. In the large scale limit, where $S_n \gg S_m$, the correlators 
$\avg{\delta_m^3}$ and $\avg{\delta_m^2 \delta_n}$ are small compared to $\avg{\delta_n^3}$ and $\avg{\delta_n^2 \delta_m}$, and
so the corresponding terms in Equation \ref{eq:fcoll_png} are only significant close to the minimum mass scale, $M_{\rm min}$ ($S_n$). 
Furthermore,
note that in the very large scale limit, $\avg{\delta_n^2 \delta_m}/S_m \propto \sqrt{1/S_m} \rightarrow \infty$, i.e., this term diverges towards large scales.\footnote{We have assumed the Newtonian form of the gravitational potential and so our calculation breaks down on scales near the horizon, where relativistic effects
are important. One might worry that this divergence is an artifact of assuming the Newtonian form for the gravitational potential.
Wands \& Slosar \cite{Wands:2009ex}, however, carry out a relativistic analysis and find that the halo bias (and hence the conditional
collapse fraction considered here) indeed diverges on large scales.} This divergence is not a concern since the density contrast, $\delta_m$, is itself tending towards zero on large scales. The $\avg{\delta_n^2 \delta_m}$ correlator in the expression here 
describes the mode-coupling between large and small scales and it is this term that 
ultimately leads to the scale-dependent halo clustering signature (e.g., \cite{Adshead:2012hs}). Presently, we are interested in the
impact of this term on the reionization barrier. Since it causes the conditional collapse fraction to blow-up on large smoothing
scales (for positive $f_{\rm NL}$), it leads to a down-turn in the reionization barrier at small $S_m$. In the case of negative
$f_{\rm NL}$, the sign of this effect is reversed and the reionization barrier turns-up at small $S_m$. We caution that approximations made in Equation \ref{eq:fcoll_png} may influence the shape of the barrier on the largest scales here,
and so we caution against taking this too literally. The behavior of
the barrier on these scales does not, however, impact our results.

\begin{figure}[htpb]
\bc
\includegraphics[width=1.0\columnwidth]{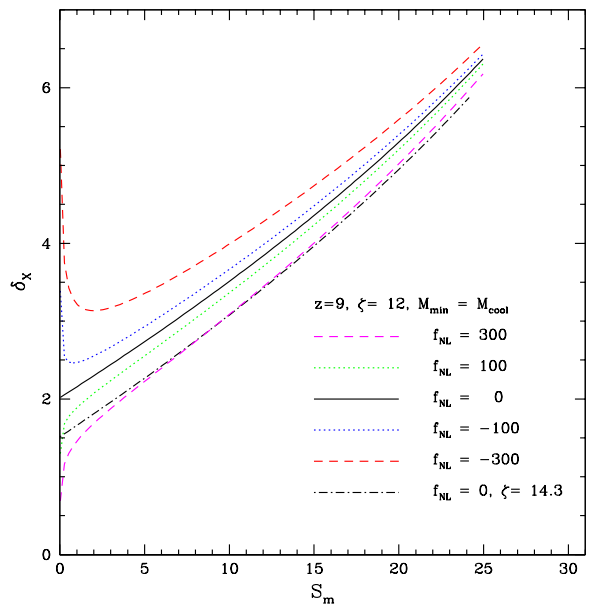}
\caption{FZH reionization barrier for various $f_{\rm NL}$ models.
The lines show the critical overdensity for a region to be ionized
as a function of the linear variance when
the density field is smoothed on the scale of the region (with both quantities linearly extrapolated to $z=0$).  Each curve
is for $z=9$, and assumes that galaxies form in halos above the atomic
cooling mass, $M_{\rm cool}$, with an ionizing efficiency parameter of
$\zeta=12$. For $f_{\rm NL}=0$, the volume-averaged ionization fraction 
is $\avg{x_i}=0.48$ at
this redshift. The collapse fraction in a region of large-scale overdensity
increases with increasing $f_{\rm NL}$ and so the critical overdensity
for a region to be ionized, $\delta_X$ {\em decreases} with increasing
$f_{\rm NL}$. The black dot-dashed line is a $f_{\rm NL} = 0$ model with $\zeta$ adjusted to match the
volume-averaged ionization fraction in the $f_{\rm NL} = 300$ model (see text).}
\label{fig:fzhbarr_v_fnl}
\ec
\end{figure}

In order to quantify the impact of $f_{\rm NL}$ on the shape of the reionization barrier, we invert
Equations \ref{eq:deltax} and \ref{eq:fcoll_png} to find $\delta_X$ as a function of $S_m$ for
several $f_{\rm NL}$ models. Here we generally consider $z=9$, $\zeta=12$, and
set $M_{\rm min}$ to
the atomic cooling mass corresponding to a virial temperature of $T_{\rm vir} = 10^4 K$, as
specified by Equation \ref{eq:mcool}. For $f_{\rm NL} = 0$ these parameters give $\avg{x_i} = 0.48$,
and so we are considering roughly the `mid-point' of reionization, where half of the volume of
the Universe is ionized. The results of these calculations are shown in
Figure \ref{fig:fzhbarr_v_fnl} for models with $f_{\rm NL} = (-300, -100, 0, 100, 300)$. The barrier
{\em decreases} with increasing $f_{\rm NL}$: positive $f_{\rm NL}$ boosts the conditional collapse
fraction in overdense regions, and hence a lower critical overdensity $\delta_X$ is required for
a region to be ionized in a non-gaussian model. The barrier turns down (up) significantly on large scales for
positive (negative) $f_{\rm NL}$ because the collapse fraction blows up towards large scales, as mentioned earlier.
Even for values as large as $|f_{\rm NL}| \sim 100$ -- now strongly disfavored by the Planck constraints~\cite{Ade:2013ydc} -- this down-turn (up-turn) occurs on rather large scales, where the barrier-crossing probability should be very 
small.
In general, $f_{\rm NL}$ leads to only small
changes in the barrier height and shape. For instance, $\delta_X$ is $\sim 8\%$ smaller at $R = 3$ Mpc/$h$ ($S_m \approx 3$) and 
$\sim 15 \%$ smaller 
at $R = 10$ Mpc/$h$ ($S_m \approx 0.5$) in the $f_{\rm NL} = 100$ model compared to the gaussian, $f_{\rm NL} = 0$, case. The 
scale $R = 3$ Mpc/$h$ mentioned
here corresponds to the peak in the analytic bubble size distribution for the $f_{\rm NL} = 0$ model, computed as in \cite{Furlanetto:2004nh},
while $\sim 99\%$ of bubbles in this model have $R \leq 10$ Mpc/$h$, the second scale considered here. These numbers hence give
some indication of which scales typically cross the reionization barrier in these models.   

In addition, the impact of $f_{\rm NL}$ on the reionization barrier at a given redshift is largely degenerate with
the effect of varying $\zeta$. This is illustrated by the black dot-dashed line in Figure \ref{fig:fzhbarr_v_fnl},
which shows an $f_{\rm NL} = 0$ model with $\zeta$ enhanced (from $\zeta =12$ to $\zeta = 14.3$) to match the 
volume-averaged ionization fraction in the
$f_{\rm NL} = 300$ model at this redshift. The shapes of the barriers are somewhat different; the barrier in the $f_{\rm NL}$ model
has the large scale down-turn and rises more steeply on small smoothing scales. The non-gaussian model should have slightly more
large bubbles and slightly fewer small bubbles than a gaussian model with the same volume-averaged ionization fraction (see \S\ref{sec:sim}
for further details).
 However, from a similar bubble-size distribution
calculation to the one mentioned above, we expect $\sim 99\%$ of random walks in the $f_{\rm NL} = 0$, $\zeta=14.3$ model (that
is largely degenerate with the $f_{\rm NL} = 300$ model) to cross the barrier at $S_m \gtrsim 0.4$. This suggests that the most prominent
differences between the two barriers, on large smoothing scales, will have little impact on the resulting bubble-size distributions since
few random walks cross the barrier on such large scales, at least near the middle of reionization. Hence it seems that varying $\zeta$ should
largely compensate for $f_{\rm NL}$ induced changes
in the reionization barrier. In addition, we should keep in mind that the values of $f_{\rm NL}$ 
considered here are already strongly disfavored by existing data, and so the degeneracy is
even more important than in this illustrative case.

\begin{figure}[htpb]
\bc
\includegraphics[width=1.0\columnwidth]{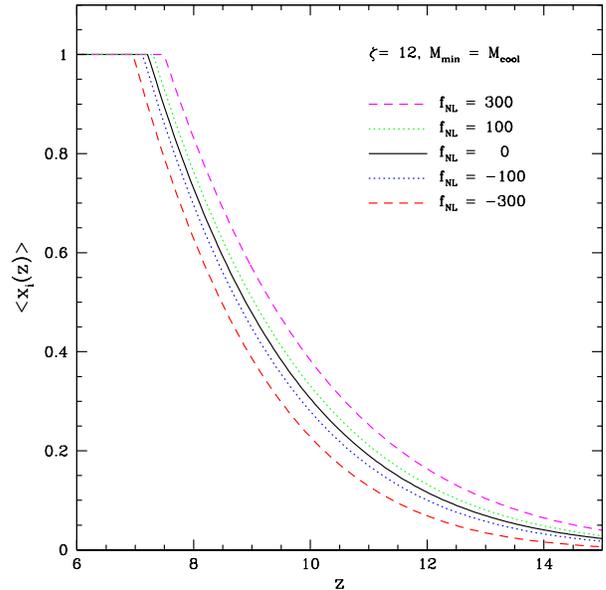}
\caption{Volume-averaged ionization fraction vs. redshift for various $f_{\rm NL}$ models. The curves show the impact of $f_{\rm NL}$ on the reionization
history of the Universe in models with $M_{\rm min}=M_{\rm cool}$ and
$\zeta=12$. The reionization process is accelerated in a Universe with
positive $f_{\rm NL}$, while it is delayed in a Universe with negative
$f_{\rm NL}$.}
\label{fig:xofz}
\ec
\end{figure}

A similar calculation determines the volume-averaged ionization fraction as a function of redshift in the FZH
model.  
Specifically, this is given by:
\begin{align}
& \avg{x_i}=\zeta f_{\rm coll}(S_n) \nonumber \\
& = \zeta {\rm erfc}\left[\frac{\delta_c(z)}{\sqrt{2 S_n}}\right] 
+ \zeta \frac{\avg{\delta_n^3}}{3 \sqrt{2 \pi} S_n^{3/2}} \left[\frac{\delta_c^2(z)}{S_n}- 1\right] {\rm exp}\left[-\frac{\delta_c^2(z)}{2 S_n}\right].
\label{eq:xav}
\end{align}
This equation holds for redshifts above which the ionized fraction, $\avg{x_i}$, becomes unity.
Here $f_{\rm coll}(S_n)$ is the (non-conditional) collapse fraction for
halos above the minimum mass at any given redshift. This is specified by Equation \ref{eq:fcoll_png} in the
limit that $M_m \rightarrow \infty$, $\delta_m \rightarrow 0$. 
Since the collapse fraction is enhanced
in models with positive $f_{\rm NL}$, so is the volume-averaged ionization
fraction. This is quantified in Figure \ref{fig:xofz}, which shows the volume-averaged ionization fraction as a function of redshift in the $f_{\rm NL}$ models of Figure \ref{fig:fzhbarr_v_fnl}.
As before, we fix $\zeta=12$ in each model and the minimum host halo
mass at the atomic cooling mass. Reionization starts and finishes
earlier (later) in models with positive (negative) $f_{\rm NL}$ compared to models 
with gaussian initial conditions. However the effects are small: for instance, the ionization
fraction near $\avg{x_i} = 0.5$ is boosted by $\lesssim 10\%$ for $f_{\rm NL} = 100$.

\begin{figure}[htpb]
\bc
\includegraphics[width=1.0\columnwidth]{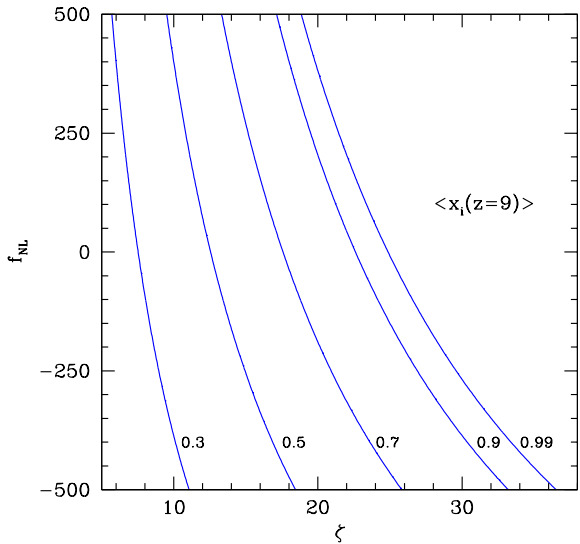}
\caption{Contours of constant ionized fraction in the $\zeta-f_{\rm NL}$ plane. The blue lines are lines of constant ionized fraction,
with $\avg{x_i(z=9)} = 0.3, 0.5, 0.7, 0.9$ and $0.99$ respectively, illustrating the degeneracy between $f_{\rm NL}$ and $\zeta$.}
\label{fig:contour_xi}
\ec
\end{figure}

The impact of $f_{\rm NL}$ on the volume-averaged ionization fraction is, however, degenerate with uncertainties
in the
ionizing efficiency and the minimum host halo mass, which are unknown and in any case 
provide only a rough model for $\avg{x_i}(z)$. We illustrate this degeneracy in Figure \ref{fig:contour_xi}, which shows contours
of constant $\avg{x_i(z=9)}$. The plot spans a very large range in $f_{\rm NL}$, including models that are already strongly disfavored
by existing data for illustration. The steepness of the contours indicates that relatively small variations in $\zeta$ can compensate
for $f_{\rm NL}$-induced changes, even over the large range in $f_{\rm NL}$ shown. In Figure \ref{fig:fzhbarr_v_fnl} we found that
the reionization barrier in an $f_{\rm NL}$ model nearly matches that in a gaussian model, once we adjust $\zeta$ to fix $\avg{x_i}$ across
models. This suggests that the bubble size distribution will mostly share this degeneracy: in other words, the contours of fixed
$\avg{x_i}$ in Figure \ref{fig:contour_xi} should resemble contours of fixed bubble-size distribution as well.
We will explore more distinctive and unique
imprints of primordial non-gaussianity subsequently.

\section{Semi-Numeric Simulations and Non-gaussianity}\label{sec:sim}

In order to calculate the statistics of the ionized regions, such as their size distribution, we need
to consider the probability that trajectories cross the barrier of Equation \ref{eq:deltax} at various
smoothing scales.
Here we will handle this numerically, using the so-called `semi-numeric' simulation technique
developed in \cite{Zahn:2006sg}.
This simulation technique is essentially a (three dimensional) Monte-Carlo implementation of
the FZH model. The Monte-Carlo implementation has the 
advantage that it partly captures asphericity
in the shapes of the ionized regions, (which we will sometimes refer to as `ionized bubbles'). Furthermore, it provides mock reionization data cubes that
are convenient for measuring the statistical properties of the epoch of reionization, and for visualizations. In
comparison to radiative transfer simulations of reionization, the semi-numeric~\cite{Zahn:2006sg} technique has
the advantage that it is extremely fast, while still capturing the large scale features of the
reionization process fairly accurately~\cite{Zahn:2006sg,Zahn:2010yw}.

In order to use this technique to study reionization in an $f_{\rm NL}$ model, we need
to first generate a non-gaussian realization of the linear density field in the model of interest.
We do this in the usual manner, briefly described here for completeness.
Specifically, we start by generating a gaussian random realization of the gaussian
field $\phi_G(\x)$. In generating this realization, we assume that
$\phi_G(\x)$ has a scale invariant power spectrum of the form $\Delta^2_\phi(k) = k^3 P_\phi(k)/(2 \pi^2) = 8.71 \times 10^{-10}$.\footnote{For reference, this model has $\sigma_8 = 0.86$, broadly
consistent with recent constraints, e.g. \cite{Bennett:2012fp}.} From $\phi_G(\x)$, we generate the primordial curvature perturbation assuming
the local model described by Equation \ref{eq:png_potential}.
The density field then follows from the potential perturbation by Poisson's Equation, which may be written (and applied) in Fourier space as:
\beqa
\delta(\k,z) = \frac{2 c^2 k^2 T(k) D(z) \Phi(\k)}{3 H_0^2 \Omega_{\rm m}}. 
\label{eq:deltaz}
\eeqa
Finally, we Fourier-transform the resulting density field into real space.  
In each model considered, the same set of random numbers are used to generate the underlying gaussian
random potential field, $\phi_G(\x)$. This lessens the impact of sample variance when comparing different
models, and isolates the impact of primordial non-gaussianity.

We generate models with $f_{\rm NL} = -100, -50, 0, 50, 100, 300$ in two different simulation volumes. The first simulation volume has a co-moving
side length of $L_{\rm box} = 150$ Mpc/$h$, while the second simulation has $L_{\rm box} = 2$ Gpc/$h$. In each case, the
density and ionization fields are tabulated on a $512^3$ Cartesian grid. The smaller simulation box captures small reionization
bubbles, while the larger volume runs are essential for examining the large scale clustering of the ionized regions, especially
the distinctive scale-dependent signatures induced by primordial non-gaussianity.

In order to construct the ionization field from a realization of the linear density field, we use the procedure of FZH and \cite{Zahn:2006sg}. 
We smooth the linear density field on a range of scales, starting from large scales and gradually stepping down to the size of the simulation
pixels. A pixel is marked as ionized if it crosses the barrier of Equations \ref{eq:deltax} and \ref{eq:fcoll_png} on some smoothing scale, while pixels that fail
to cross the barrier on any smoothing scale are marked neutral. Since we have proper non-gaussian realizations of the density field, the enhanced probability of 
crossing the
reionization barrier in $f_{\rm NL}$ models is naturally accounted for in this step.

\begin{figure}[htpb]
\bc
\includegraphics[width=1.0\columnwidth]{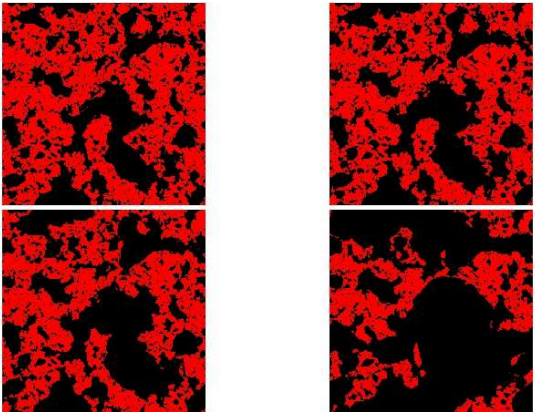}
\caption{Slices through semi-numeric simulations of reionization 
models with primordial non-gaussianity. In each model, $\zeta=12, z=8$. The
slices are $150$ co-moving Mpc/$h$ on a side, and are
$0.29$ Mpc/$h$ thick.  
{\em Top Left}: $f_{\rm NL}=-100$, {\em Top Right}: $f_{\rm NL}=0$,
{\em Bottom Left}: $f_{\rm NL}=100$, {\em Bottom Right}: $f_{\rm NL} = 300$.
The black regions show ionized bubbles, while the red regions are
neutral gas. As $f_{\rm NL}$ increases, reionization is more progressed
(for fixed ionizing efficiency, $\zeta$, and redshift, $z$), and the
bubbles are larger.}
\label{fig:bub_examp}
\ec
\end{figure} 

We show example slices through $150$ Mpc/$h$ semi-numeric reionization simulations with $\zeta=12$ and $z=8$ in
Figure \ref{fig:bub_examp}. Since each simulation is generated with the same underlying gaussian random part of the potential field, $\phi_G(\x)$,
we can compare the slices directly, region-by-region. As expected, reionization has progressed further, and the ionized
regions are larger, in the models with primordial non-gaussianity. In the $f_{\rm NL} = \pm 100$ model, however, the differences
with the case of gaussian initial conditions are somewhat subtle. 
Note that we are comparing the different models at fixed $\zeta$ and $z$, and so the models have
varying volume-averaged ionization fractions, which increase with $f_{\rm NL}$. Specifically,
the model with $f_{\rm NL} = 0$ has $\avg{x_i} = 0.45$ at $z=8$, while
$\avg{x_i} = (0.40, 0.50, 0.63)$ for $f_{\rm NL} = (-100, 100, 300)$ at $z=8$. A caveat here is that these values are 
smaller than expected from the analytic curves in Figure \ref{fig:xofz}, and the difference with the expected value
decreases with increasing $|f_{\rm NL}|$. The smaller values result for two reasons: first, there 
are some ionized bubbles that are smaller than the size of our simulation grid and
hence not captured. This effect is more important at small $|f_{\rm NL}|$ since the bubbles are smaller in these models. 
Second, our collapse-fraction expressions are derived using a top-hat filter in $k$-space but applied with
a real-space filter: as discussed in the Appendix of \cite{Zahn:2006sg}, this leads to departures from the expected global
ionized fraction. Since the departure decreases with increasing $f_{\rm NL}$, our semi-numeric simulations 
likely overestimate the impact of $f_{\rm NL}$ on
the ionized fractions and bubble-sizes, but they still serve to illustrate the main effects of non-gaussianity on reionization.

\begin{figure}
\bc
\includegraphics[width=1.0\columnwidth]{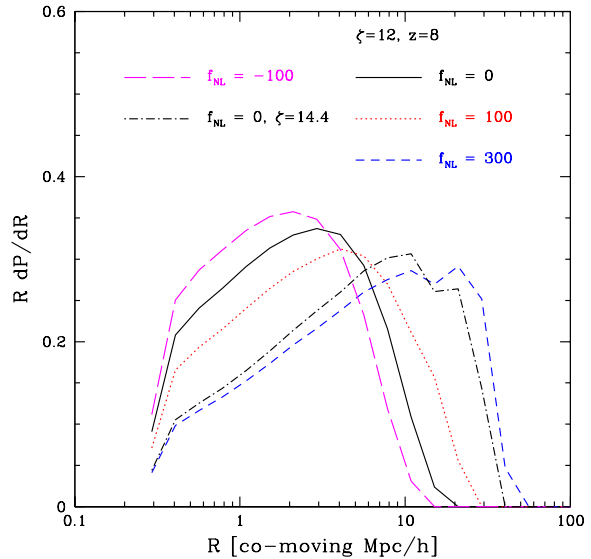}
\caption{Bubble-size distribution for various $f_{\rm NL}$ models. The
curves show the probability distribution of the sizes of the ionized
regions for $x_{\rm th}=0.9$ (see text). The models are at $z=8$ and most assume $\zeta=12$, $M_{\rm min} = M_{\rm cool}$, for
various values of $f_{\rm NL}$ as labeled. 
The black dot-dashed model is for an $f_{\rm NL} = 0$ model with $\zeta$ adjusted to match the mean ionized
fraction ($\avg{x_i}$) in the $f_{\rm NL} = 300$ model. This illustrates that the variations in bubble size with $f_{\rm NL}$ 
at fixed $\avg{x_i}$ are much smaller than at fixed $\zeta$.
}
\label{fig:bubble_pdf}
\ec
\end{figure}

In order to quantify the visual impressions of Figure \ref{fig:bub_examp},
we calculate the probability distribution of the sizes of the ionized regions for each
$f_{\rm NL}$ model. This depends somewhat on one's definition for the size of
the complex, aspherical ionized regions in the simulation.
Here we define the size of the simulated ionized
regions as in \cite{Zahn:2006sg}. Briefly, we spherically average the ionization
field on various scales $R$, starting from large scales, stepping downward
in size until we eventually get to the size of our simulation pixels. At
each smoothing scale, we compare the spherically averaged ionization field
to a threshold ionization value, $x_{\rm th}$. A simulation pixel is considered
`ionized' and belonging to a bubble of radius $R$, when $R$ is the largest
smoothing radius at which the smoothed ionization field crosses the threshold.
Pixels that do not cross the threshold on any smoothing scale are considered
neutral. This procedure is of course similar to the way in which the ionization field is constructed
in the first place.

The resulting probability distribution function (PDF), for $x_{\rm th}
= 0.9$, is shown for our various $f_{\rm NL}$ models with $\zeta=12$,
$z=8$ in Figure \ref{fig:bubble_pdf}. Note that each PDF is normalized
to $1$ rather than to the ionized fraction: the PDFs show the fraction
of bubbles that have a radius between $R$ and $R + dR$.  To mention
one quantitative description of these results, we examine by how much
the characteristic size of ionized bubbles -- which we identify with
the peak in the bubble size PDFs of Figure \ref{fig:bubble_pdf} --
varies with $f_{\rm NL}$ for the present values of $\zeta,
z$. Compared to a model with gaussian initial conditions, the
characteristic bubble size increases by a factor of $1.4$ for $f_{\rm
  NL} = 100$, a factor of $3.7$ for $f_{\rm NL} = 300$, while it
decreases by a factor of $1.5$ for $f_{\rm NL} = -100$. As anticipated
in the previous section, the bubble-sized distribution is degenerate
with $\zeta$. The black dot-dashed line shows that an $f_{\rm NL} = 0$
model with $\zeta$ adjusted upward to match the ionized fraction in
the $f_{\rm NL} = 300$ model. The resulting bubble-size distributions
are quite similar, although the bubbles are a little bit larger in the
$f_{\rm NL}$ model. This is expected from Figure
\ref{fig:fzhbarr_v_fnl}; the main difference between the $f_{\rm NL}$
barrier and the gaussian barrier at fixed $\avg{x_i}$ is that the
$f_{\rm NL}$ barrier is a little lower on larger scales (smaller
$S_m$) allowing slightly larger ionized regions to form.

One might wonder whether random walks can start to cross the reionization barrier at very low $S_m$ in non-gaussian models, where
the barrier has this distinctive downturn. In principle, this might lead to a 
bi-modality in the bubble size distribution: perhaps the downturn allows some trajectories to cross the barrier on very large scales that would be prohibited from crossing otherwise. On smaller smoothing scales, the trajectory-crossing probability might shrink as the smoothing scale becomes smaller than the down-turn scale and the barrier increases, until trajectories catch up again close to 
the scale of the usual peak
in the bubble size distribution. This might imprint a distinctive large-scale bump in
the bubble size distribution. It is possible
that this is even the origin of the kink in the high-$R$ tail of the $f_{\rm NL} = 300$ model in Figure \ref{fig:bubble_pdf}. This effect might be more realizable
at the end of reionization, when random walks start crossing the reionization barrier on progressively larger scales. However, this effect
is unlikely important for the much smaller values of $f_{\rm NL}$ presently allowed, and so we don't investigate it further here. It also may be an artifact
of approximations made in deriving Equation \ref{eq:fcoll_png}, as mentioned
earlier.

\section{Scale-dependent Bias}
\label{sec:scale_b}

The previous results describe the general impact of primordial
non-gaussianity on reionization. However, the most promising approach
for obtaining observational constraints on non-gaussianity from
reionization studies is to measure the scale dependent clustering of
the ionized regions. This is directly analogous to the case of the
clustering of dark matter halos. In the case of halo clustering, Dalal
et al.~\cite{Dalal:2007cu} showed that local non-gaussian models give
rise to a scale-dependent clustering signature. If a similar signature
arises in the clustering of ionized regions, this may provide a
distinctive indicator and allow constraints on $f_{\rm NL}$ from the
epoch of reionization, despite uncertainties in the properties of the
ionizing sources, and the surrounding IGM.

Indeed, we would expect the ionized regions to have a similar scale-dependent clustering signature to that of the dark matter halos. In terms of excursion set modeling, the reionization case is different than halo clustering only in that
the reionization barrier (Equation \ref{eq:deltax}) has a different shape than the halo collapse barrier. Otherwise the physics underlying the scale dependent
clustering is similar. In particular,
a positive value of $f_{\rm NL}$ implies that the {\em variance} of the density
field is enhanced in large scale regions with above average potential 
perturbation (which are overdense on large scales as quantified by the Poisson Equation, Equation \ref{eq:deltaz}.) The enhanced variance boosts the
collapse fraction in such regions, and {\em increases} the tendency for these regions to 
be ionized before typical regions. In the context of the FZH model, the extra
variance in large scale overdense regions boosts the rate at which trajectories
cross the barrier of Equation \ref{eq:deltax} just as it increases the rate of crossing
the spherical collapse barrier in the case of halo clustering.  
In both cases the form of
Equation \ref{eq:deltaz} then implies a distinctive scale-dependent clustering term,
$\propto 1/(k^2 T(k))$~\cite{Dalal:2007cu}. Moreover, we~\cite{Adshead:2012hs} (see also \cite{DAloisiobias})
showed that this general form is expected for an arbitrary barrier crossing problem in the presence of primordial non-gaussianity with a non-zero squeezed limit. 
One caveat in the reionization case is that we expect this to apply only on scales much larger than
the size of the ionized regions. On smaller scales, the ionization field will decorrelate from the underlying density field 
since small scale regions are either highly ionized or completely neutral irrespective
of the precise value of the density field. 

Our aim here is to quantify the scale-dependent clustering of the ionized regions. In order to
capture the large scale Fourier modes where the scale-dependent clustering signature should dominate,
we use $2$ Gpc/$h$ semi-numeric reionization simulations. 
As before, we construct simulations for several values of $f_{\rm NL}$; here we
simulate $f_{\rm NL} = 0, \pm 50, \pm 100$.  We further calculate the power spectrum of the ionization field in each model, $P_{x,x}(k)$.\footnote{We consider
the power spectrum of $x_i$ rather than $\delta_x = (x_i - \avg{x_i})/\avg{x_i}$, i.e., we do not normalize by
$\avg{x_i}$.} We also calculate the ionization-density cross power spectrum, $P_{x,\delta_\rho}(k)$ and the density power spectrum, $P_{\delta_\rho,\delta_\rho}(k)$. We then estimate the bias of the ionization field from the ratio of the 
ionization-density cross power spectrum and the density power spectrum,
\beqa
b_x(k) = P_{x,\delta_\rho}(k)/P_{\delta_\rho,\delta_\rho}(k).
\label{eq:biasx}
\eeqa
 
\begin{figure}
\bc
\includegraphics[width=1.0\columnwidth]{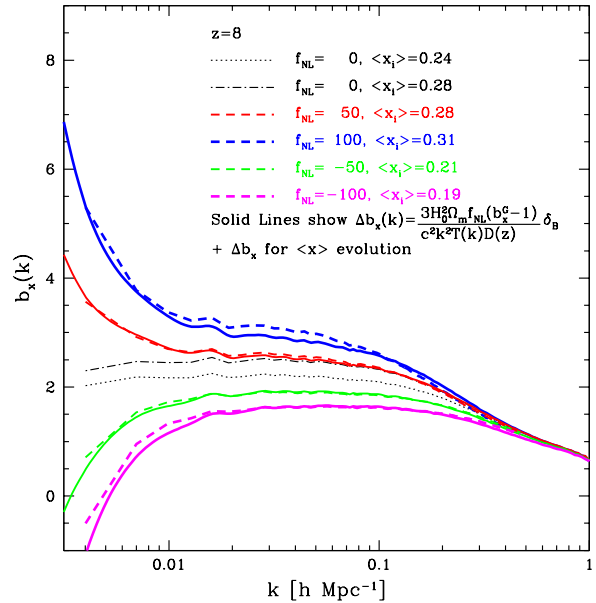}
\caption{Scale-dependent bias of ionization field. The dashed lines show measurements of $b_x(k)$ (Equation \ref{eq:biasx}) from semi-numeric
simulations for $z=8$ and fixed $\zeta=12$. The solid lines show (by eye) fits to the simulation results. The fits consist of two terms. The first term is a 
scale dependent
enhancement in the form of Equation \ref{eq:scale_bias}, with the proportionality constant, $\delta_B$, adjusted to (roughly) match the $f_{\rm NL} = 50$ measurement.
The second term is predicted by varying $\zeta$ in a gaussian model to match the $\avg{x_i}$ value in the corresponding $f_{\rm NL}$ model. As an illustrative example,
the dot-dashed line shows $b_x(k)$ for an $f_{\rm NL} = 0$ model with the same $\avg{x_i}$ as for $f_{\rm NL} = 50$. The difference between the
dashed and dot-dashed lines hence reflects the roughly scale-independent (on large scales) correction to the bias for $f_{\rm NL} = 50$, while
the enhancement at $k \lesssim 0.02 h$ Mpc$^{-1}$ is dominated by the scale-dependent contribution. 
}
\label{fig:bofx}
\ec
\end{figure}

As discussed previously, we expect primordial non-gaussianity to slightly lower the collapse barrier of
Equation \ref{eq:deltax} (for positive $f_{\rm NL}$, as illustrated in Figure \ref{fig:fzhbarr_v_fnl}), and 
for the enhanced high density tail  
in these models to increase the probability of crossing the reionization barrier. The latter
effect, and the coupling between large and small scale modes, is responsible for the scale dependent clustering
enhancements. However, each of these two effects modifies the bias of the ionized regions, $b_x(k)$. Since $f_{\rm NL}$ is known to be small, we can consider
each effect as a small correction in a Taylor expansion around $f_{\rm NL}=0$,
and calculate the two effects separately. 
The change in bias from
the reduced barrier height in a positive $f_{\rm NL}$ model is mainly due to the fact that the ionization fraction is larger
(than gaussian). For small changes in $f_{\rm NL}$, and on scales much larger than the ionized
regions, this leads to a roughly scale-independent change in $b_x(k)$. 
We find that this scale-independent term is fairly well matched by considering 
the change in bias in an $f_{\rm NL} = 0$ model after adjusting
$\zeta$ to match the enhanced $\avg{x_i}$ in 
the corresponding $f_{\rm NL}$ model (see Figure \ref{fig:bofx} for an illustration). 

The second effect -- increased probability of barrier crossing in a non-gaussian model -- 
produces a distinctive scale-dependent clustering enhancement
of the \cite{Dalal:2007cu} form:
\beqa
\Delta b_x(k) =  \frac{3 H^2_0 \Omega_{\rm m} f_{\rm NL} (b_x^G - 1)}{c^2 k^2 T(k) D(z)} \delta_B.
\label{eq:scale_bias}
\eeqa
Here $\Delta b_x(k)$ denotes the change in bias with wavenumber $k$ owing to $f_{\rm NL}$, and the above equation describes only
the scale dependent contribution. The quantity $b_x^G$ is the bias of the ionization field in a gaussian model, and $T(k)$ is the
transfer function. Here $\delta_B$ is a proportionality constant related to the height of the reionization barrier. Note that $b_x^G$ and $\delta_B$ depend
on both redshift/ionization fraction and the reionization model.
The growth factor
here, $D(z)$, is the linear growth factor normalized to $1/(1+z)$ during the matter-dominated era.

Our numerical results mostly perform this perturbative analysis, as shown in Figure
\ref{fig:bofx} for $z=8$ and $\zeta=12$.\footnote{The ionized fraction is lower by a factor of $\sim 2$ than in our $150$ Mpc/$h$ volume because some of the small 
bubbles are unresolved in the large volume simulated here, which focuses on capturing the large scale bias. The limited resolution
of our calculations should impact the bias numbers a little at a given $\avg{x_i}$, but should not impact the main trends.}
We adjust the single parameter $\delta_B$ in Equation \ref{eq:scale_bias} above
to match the simulation results,
and calculate the scale-independent correction to the bias as described
above. The result is shown by the solid curves in Figure \ref{fig:bofx}.
The fit is a fairly good overall match to the simulation results, although
some differences are evident.
Most importantly, one can see the scale-dependent enhancement of
the Dalal et al.~\cite{Dalal:2007cu} type on large scales, and that this term is linear in $f_{\rm NL}$, as expected. (Note that the `fit' curves share noisy features with the simulation results, because the
scale independent correction is calculated directly from the simulated power spectra.)

The fits, however, seem 
to slightly underproduce the
large scale enhancement for $f_{\rm NL} = 100$, and slightly overproduce the effect at $f_{\rm NL} = -100$. The departure from the fits at large $f_{\rm NL}$
likely results because the change in the scale-independent bias in these models
is becoming significant and is not perfectly captured by matching to a fixed $\avg{x_i}$. After all, the reionization barrier and bubble size distribution still
differ slightly after matching to a fixed $\avg{x_i}$ (see Figure \ref{fig:fzhbarr_v_fnl}). 
Along these lines, we find a better fit if, in calculating the scale-independent correction, 
we boost $\zeta$ a bit beyond that required to match
the enhanced $\avg{x_i}$ in the non-Gaussian model (although this improvement is not shown in the figure here).
Quantitatively, we find $(b_x^G -1) \delta_B \approx 0.49$ for the redshift
and ionization fraction considered here. This coefficient also depends
on the particular reionization model considered here, e.g., the assumption
that the ionizing efficiency in Equation \ref{eq:mion} is itself independent
of mass.
Note that we considered the scale-dependent bias of $x_i$ here, if we had instead considered
the field $\delta_x = (x_i - \avg{x_i})/\avg{x_i}$ the scale dependent bias coefficient would
be $\approx 2$.

The relatively good match emboldens us to use the simple fitting
formula of Equation \ref{eq:scale_bias} in \S \ref{sec:observability} to project how well future 21 cm surveys will be able to extract $f_{\rm NL}$. The
imperfect fit does, however, suggest that more theoretical work will be
needed to extract the exact value of $f_{\rm NL}$ from the data.\footnote{As we were finishing this work, Ref. \cite{D'Aloisio:2013sda} was posted, making essentially
the same point; quantitatively their results may differ a bit more
from Equation \ref{eq:scale_bias} than do ours, but the qualitative conclusion
of a scale-dependent bias has now been verified by several different groups
using a range of techniques.}

Note that the scale dependent enhancement dominates
over the variation from the enhanced $\avg{x}$ only on very large scales (small $k$), $k \lesssim 0.02 h$ Mpc$^{-1}$. Unfortunately, these scales will likely be challenging to 
observe (see \S\ref{sec:observability}). In practice, the efficiency of the ionizing sources and their host halo masses will not be known {\em a priori} and
so we will likely need to perform a joint fit for the scale independent and scale dependent bias contributions. If $\avg{x_i}(z)$ can be determined
from independent observations, this would be helpful in separating out the scale dependent enhancement. 

We also investigated how the results depend on the particular stage of
the reionization process, by considering an additional model in which
$\zeta$ is enhanced beyond our fiducial values of $\zeta=12$. In the
alternate case considered, $\avg{x_i} = 0.51$ at $z=8$ (for $f_{\rm
  NL} = 0$) and so it describes the impact of non-gaussianity near
reionization's midpoint.  In this case, we find a similar fit works,
especially if we allow a slight boost to the scale-independent term,
beyond that required to match to a fixed $\avg{x_i}$. At this stage of
reionization, we find $(b_x^G -1) \delta_B \approx 0.35$. The
proportionality constant in the fit of Equation \ref{eq:scale_bias}
hence varies with $\zeta$, but this is to be expected since the
proportionality constant should depend on the barrier height, which
itself depends on $\zeta$.  In principle, one can also consider later
stages in reionization, but we expect this to be less useful. First,
the signal drops off towards the end of reionization
(e.g. \cite{Lidz:2007az}).  Second, note that the ionization and
density fields decorrelate on scales less than the size of the ionized
regions. As the ionized bubbles grow, only larger and larger scales
are useful for the scale-dependent clustering signature.

It is also possible to compute the scale-dependent bias coefficient,
$(b_x^G -1) \delta_B$, analytically. The analytic calculations allow one to quickly explore many $f_{\rm NL}$ models for a wide range of ionized-fractions, $\avg{x_i}$, and redshifts.
Ref. \cite{Adshead:2012hs} has
derived expressions for the scale-independent bias and scale-dependent
bias coefficient that are applicable to any collapse barrier using a
path integral approach developed by \cite{Maggiore:2010I,
  Maggiore:2010II, Maggiore:2010III}.  
Following
\cite{Adshead:2012hs}, we define $\mathcal{P}_0(S_m)$ as an expansion
in the barrier:
\begin{eqnarray}
\mathcal{P}_{0}(S_m) = \sum_{p = 1}^5 \frac{-S_m^p}{p!} \frac{d^p}{d S_m^p}  \delta_X(S_m).
\end{eqnarray}
The scale-independent part of the bias can then be written as 
\begin{eqnarray}
b_x^G(S_m) = 1 + \frac{\delta_X(S_m)D(0)}{D(z)S_m} - \frac{D(0)}{D(z)(\delta_X(S_m) + P_0(S_m))} \nonumber \\
\end{eqnarray}
which reduces to the standard expression from e.g. \cite{McQuinn:2005} in the limit that the barrier is linear. Since $\delta_X(S_m)$ and $S_m$ are linearly extrapolated to $z=0$, the growth factor enters here in calculating the bias
at redshift $z$.  
For an $f_{\rm NL}$ cosmology the  scale-dependent bias can be written \cite{Adshead:2012hs}
\begin{align}\label{scaledepbiasgen}
b^{\rm SD}(k) = \frac{2f_{\rm NL}}{\mathcal{M}(k)} c(S_m)  
\end{align}
where the coefficient $c(S_m)$ is defined
\begin{eqnarray}
\label{eq:sdcoeff}
c(S_m) &=& \delta_X(S_m) (b_x^G-1) \frac{D(z)}{D(0)} -\frac{3\mathcal{P}_{0}(S_m)}{\delta_X(S_m)+\mathcal{P}_{0}(S_m) } \nonumber \\ 
&& + \left(2 \sqrt{\frac{2\pi}{S_m}}\frac{\mathcal{P}^2_{0}(S_m) }{(\delta_X(S_m)+\mathcal{P}_{0}(S_m))}\right) \nonumber \\
&& \times \text{exp}\left[\frac{\delta_X^2(S_m)}{2 S_m}\right] \text{erfc}\left[\frac{\delta_X(S_m)}{\sqrt{2S_m}}\right].
\end{eqnarray}
and
\begin{align}
\mathcal{M}(k) = \frac{2}{3}\frac{c^{2}k^{2}T(k)D(z)}{\Omega_m H_0^2}
\end{align}
In the limit that barrier is flat, i.e. $\delta_X(S_m) = \delta_c$, then $\mathcal{P}_0(S_m) = 0$ and the above expression reduces to
the standard prediction $c(S_m) = \delta_c(b_x^G -1 )D(z)/D(0)$, which has
been derived in e.g. \cite{Dalal:2007cu}.

To compare these analytic expressions for the bias with the results
of the semi-numeric simulations presented above, we must calculate the
volume-weighted average of $b_x^G(S_m)$ and $c(S_m)$ over all ionized
bubbles.  This averaging is accomplished by integrating over the
bubble size distribution:
\begin{eqnarray}
\bar{b}_x^G &=& \avg{x_i}^{-1} \int dm \, b_x^G(S(m)) \frac{dn}{dm} V(m) \\ 
\bar{c} &=& \avg{x_i}^{-1} \int dm \, c(S(m)) \frac{dn}{dm} V(m)
\end{eqnarray}
where 
\begin{eqnarray}
\avg{x_i} = \int dm \frac{dn}{dm} V(m)
\end{eqnarray}
and $V(m) = m/\bar{\rho}$ is the volume of a region of mass $m$.  The mass function $dn/dm$ can be computed analytically from the excursion set formalism:
\begin{eqnarray}
\frac{dn}{dm} = \frac{1}{m} \frac{\bar{\rho}}{m} \left| \frac{d \ln S(m)}{d \ln m}\right| S(m) \mathcal{F}(S(m))
\end{eqnarray}
where $\mathcal{F}$ is the so-called unconditioned crossing rate
discussed in \cite{Adshead:2012hs}.

\begin{figure}
\includegraphics[width=1.0\columnwidth]{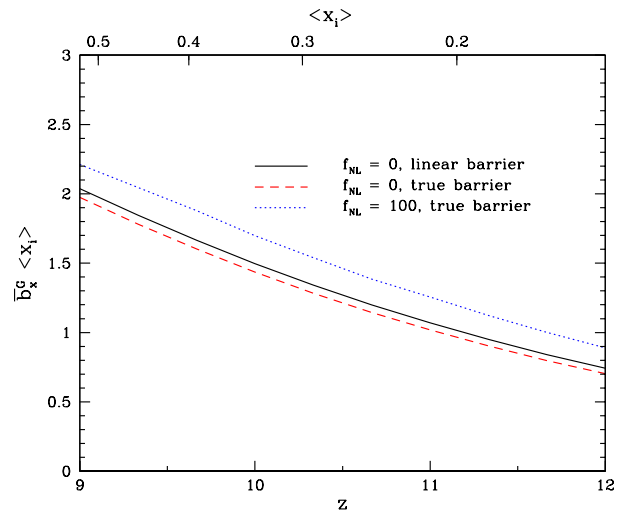}
\caption{The volume averaged scale-independent bias. }
\label{fig:mean_si_bias}
\end{figure}

\begin{figure}
\includegraphics[width=1.0\columnwidth]{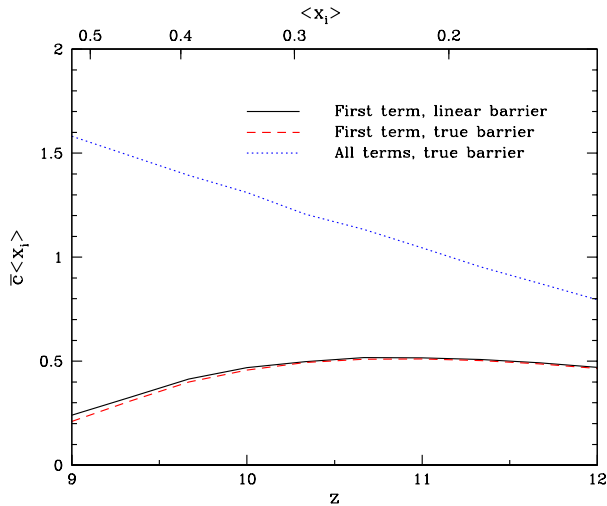}
\caption{The volume averaged coefficient of the scale-dependent bias. }
\label{fig:mean_sdcoeff}
\end{figure}

The results of our calculation of the volume averaged
scale-independent bias and volume average scale-dependent bias
coefficient are shown in Figures \ref{fig:mean_si_bias} and
\ref{fig:mean_sdcoeff}.  In these figures we have multiplied $\bar{b}_x^G$ by
$\avg{x_i(z)}$ to obtain the bias of $x_i$ rather than the field
$\delta_x = (x_i - \avg{x_i})/\avg{x_i}$; this allows
for direct comparison to the semi-numerical simulation results
presented above.  Figure \ref{fig:mean_si_bias} shows the result of
calculating the bias for three different collapse barriers: a linear
fit to the gaussian barrier (black solid curve), the true gaussian
barrier (red dashed curve), and the barrier with the effects of
non-gaussianity included at the level of $f_{\rm{NL}} = 100$ (blue
dotted curve).  The first of these barriers corresponds to the
calculation of the bias presented in \cite{Furlanetto:2004nh}.  We see
that the effects of non-gaussianity at the level of $f_{NL} = 100$ on
the mean scale-independent bias are comparable in magnitude to the
effects of approximating the barrier as a linear function of the variance.

Figure \ref{fig:mean_sdcoeff} shows three different approaches for
calculating the scale-dependent bias coefficient.  The black
(solid) and red (dashed) curves compute this
coefficient using only the first term of Equation \ref{eq:sdcoeff}
(i.e. the standard result) for a linear barrier and the true
barrier, respectively.  The blue (dotted) curve, on the other hand, includes all of the terms in Equation \ref{eq:sdcoeff}.
Including only the first term, the linear and true barrier calculations
give very similar results.
Including all of the terms, however, gives a significantly
larger scale-dependent bias; this
difference grows with decreasing redshift as reionization proceeds.
We find that the first term approximation gives better agreement with the simulation
results. In \cite{Adshead:2012hs}, we found
related differences between the first term approximation and including
all of terms for the case of
an ellipsoidal collapse barrier (see Figure 3 of \cite{Adshead:2012hs}). In that paper, it was speculated that
approximations used in deriving Eq. \ref{eq:sdcoeff} -- in particular, approximating 3-point functions by their end-point values -- may lead to an artificial
rise in the scale-dependent bias coefficient when all of the
terms in Equation \ref{eq:sdcoeff} are included. A similar effect may make this coefficient
spuriously large here.

The analytic calculations roughly agree with our numerical results, provided
the first-term approximation is more robust and so we compare with it and
adopt this approximation in what follows.
The scale-independent bias is roughly $1.5$ and the scale dependent bias
coefficient -- denoted by $\bar{c} \avg{x_i}$ in the analytic calculation of Equation \ref{eq:sdcoeff} -- is roughly $(b_x^G - 1) \delta_B \approx 0.5$ over the
range of $z$ that we consider. This is close to the simulated values at similar ionized fractions (see Figure \ref{fig:bofx}).
There are several reasons why the
analytic calculation might not agree exactly with the simulation
results.  For one, the analytic calculation assumes that the ionized
regions are spherical, while it is clear from Figure \ref{fig:bub_examp}
that this is not the case. Second, the analytic results are based on
the first-term approximation to Equation \ref{eq:sdcoeff}. In order to robustly calculate the corrections here, it may be necessary to move beyond 
approximating the
3-point functions at their end-point values, as mentioned above. Next,
the simulations here do not resolve the smallest bubbles, and do not produce
precisely the expected $\avg{x_i}$, as discussed previously. Asides
for these caveats, we find the agreement between the analytic and numerical calculations encouraging and will therefore use the analytic calculations to estimate the prospects for constraining $f_{\rm NL}$ with future surveys (\S \ref{sec:observability}).

Our semi-numeric results regarding scale-dependent biasing also agree broadly with Joudaki et al.~\cite{Joudaki:2011sv}, who first identified the scale-dependent clustering enhancement in the ionization field.
These authors approximated, however, the reionization barrier as fixed,
while we further included the impact of primordial non-gaussianity on
the reionization barrier itself.  As a result, they did not include
the scale-independent enhancement discussed here. However, we have
verified that this is not a big obstacle, and that the scale-dependent
signature can still be easily discerned. We expect this to be even
more the case for smaller values of $f_{\rm NL}$ than considered here;
given the tight constraints from Planck data, smaller values of
$f_{\rm NL} \sim \pm 10$ span the currently interesting regime.

\section{Is it Measurable?}
\label{sec:observability}

Now that we have characterized the impact of primordial
non-gaussianity on reionization and quantified its signature in the
scale-dependent clustering of ionized regions, we turn to consider
whether these signatures may be observable in upcoming reionization
surveys. The most promising approach is to use the scale-dependent
bias, since the other effects are likely degenerate with uncertainties
in the properties of the ionizing sources and the surrounding
IGM. Three main types of observations have been discussed in the
literature that can potentially probe or constrain the scale-dependent
clustering of the ionized regions: narrow-band surveys for Ly-$\alpha$
emitting galaxies during reionization
(e.g. \cite{Furlanetto:2005ir,McQuinn:2007dy}), measurements of the
small-scale CMB anisotropies induced by the patchy kinetic
Sunyaev-Zel'dovich effect (e.g. \cite{Zahn:2011vp,Mesinger:2011aa}),
and measurements of the redshifted 21 cm line from the epoch of reionization (EoR)
(e.g. \cite{Furlanetto:2006jb}). The latter measurement is the most
direct probe of spatial fluctuations in the ionized fraction during
reionization, and so we focus on these measurements here.

The main quantity of interest for the redshifted 21 cm measurements is
the 21 cm brightness temperature contrast between a neutral hydrogen
cloud at redshift $z$ and the CMB. We work in the limit that the
spin temperature of the 21 cm line is much larger than the CMB
temperature throughout all space, and further, we ignore redshift-space
distortions from peculiar velocities. These are expected to be good
approximations during most of the EoR
(e.g. \cite{Ciardi:2003hg,Mesinger:2010ne,Mao:2011xp}).\footnote{Since spin temperature fluctuations are expected to be coherent on rather large scales \cite{Mesinger:2010ne}, they may in fact make it harder to extract the signatures of primordial non-gaussianity compared to the simplified case considered here. This should mostly impact the early stages of reionization, and so we don't consider the effects of spin temperature fluctuations further here.}
With these
approximations, the brightness temperature contrast is: 
\beqa T_{\rm
  21} = T_0 x_{\rm HI} (1 + \delta_\rho).
\label{eq:t21}
\eeqa
The constant $T_0$ is $T_0 = 28 \left[(1+z)/10\right]^{1/2}$ mK for our cosmological parameters \cite{Furlanetto:2006jb}. Here $x_{\rm HI}$ is the neutral fraction, and $\delta_\rho$ is the density contrast. Each of $T_{\rm 21}$, $x_{\rm HI}$, and $\delta_\rho$ vary spatially and with redshift, but we have suppressed this dependence in our notation here.

The power spectrum of 21 cm brightness temperature fluctuations is related
to the spatial fluctuations in the ionization field we considered earlier.
If we expand to first order in fluctuations in the density contrast and
to first order in neutral fraction fluctuations (neglecting redshift space distortion and spin temperature fluctuations), we expect:
\beqa\label{eq:pk21}
\Delta^2_{\rm 21, 21}(k) &\approx& T_0^2 \left[b_x^2 - 2 (1-\avg{x_i}) b_x + (1-\avg{x_i})^2\right] \nonumber \\
&& \times \Delta^2_{\delta_\rho, \delta_\rho}(k).
\eeqa

Here $b_x$ is defined as in Equation \ref{eq:biasx}: it is the linear bias factor of the
ionization field, rather than the bias of the {\em fluctuations in the ionization field} that is sometimes considered. The bias factor we consider here is equal to the bias factor of the ionization fluctuations, multiplied by a factor of $\avg{x_i}$. In general, working to first order in $\delta_x$ can be problematic since
the ionized regions are expected to be large during most of the EoR. The large spatial fluctuations in the ionization fraction imply 
that additional terms,
dropped in Equation \ref{eq:pk21}, can be important even on large spatial scales \cite{Lidz:2006vj}. Since we are interested here only in the very large scale clustering signature, that applies on
scales much larger than the size of the ionized regions, the approximation
of Equation \ref{eq:pk21} should nonetheless be adequate. 
In this approximation,
Equation \ref{eq:pk21} relates the bias factor of the 21 cm fluctuations
to the bias factor measured in the previous section. 

A challenge for measuring the scale-dependent clustering signature
with future 21 cm measurements is that primordial non-gaussianity
significantly impacts clustering only on rather large scales. On these large scales
foreground contamination in the redshifted 21 cm data may be
prohibitive. At the frequencies of interest, foreground contamination
from galactic emission and extragalactic point sources is expected to
have a mean brightness temperature that is roughly four orders of
magnitude larger than the average redshifted 21 cm signal from the
epoch of reionization. Fortunately, the foreground contamination is
expected to be spectrally smooth and distinguishable from the
redshifted 21 cm signal which should have significant frequency
structure (e.g. \cite{McQuinn:2005hk}).  Nevertheless, cleaning this
contamination will prohibit measuring the signal for long wavelength
modes along the line of sight, and weaken the ability to measure the
large-scale power spectrum, precisely where primordial non-gaussianity
should have its most pronounced effects.

The impact of foreground cleaning depends on the bandwidth over which
the contamination is estimated, and the precise algorithm used to
separate or subtract the slowing varying line of sight modes
(e.g. \cite{McQuinn:2005hk,2011MNRAS.413.2103P}). One commonly
discussed approach is to fit a low order polynomial in frequency to
each interferometric pixel in the Fourier $(u,v)$ plane. The optimal
choice for this method is to use the lowest order polynomial for which
the residual foreground power is well beneath the signal power.
Higher order polynomials are required to fit the foregrounds if the
fitting is done over larger bandwidths. Previous work suggests that a
cubic polynomial, described by $N=4$ coefficients, is adequate for
fitting foregrounds over a bandwidth of $32$ Mhz
\cite{Bowman:2008mk}.\footnote{This full bandpass may be divided
  up into smaller chunks, of perhaps $6$ Mhz width, for power spectrum
  estimation. This is to ensure that the power spectrum evolves
  minimally across the redshift range of a chunk
  \cite{McQuinn:2005hk}.}  For reference, the co-moving length
scale corresponding to $32$ Mhz of bandwidth is \beqa L_{\rm band} =
390\, {\rm Mpc}/h \left[\frac{B}{32\, {\rm Mhz}}\right]
\left[\frac{0.27}{\Omega_{\rm m}}\right]^{1/2}
\left[\frac{1+z}{9}\right]^{1/2}. \nonumber \\
\label{eq:lband}
\eeqa
A simple estimate is that a polynomial of order $N$ has $N-1$ nodes, and hence removing a polynomial
fit of this order removes modes with line-of-sight wavenumber $|k_{\parallel, \rm min}| < N \pi/ L_{\rm band}$ \cite{2012Natur.487...70V}. This suggests that the minimum measurable wavenumber, $k^2 = k_\parallel^2 + k_\perp^2$, after foreground cleaning is
$k_{\rm min} = |k_{\parallel, \rm min}| = N \pi/L_{\rm band}$. If $N=4$ and $B=32$ Mhz, $k_{\rm min} = 0.032 h$ Mpc$^{-1}$. Comparing with
Figure \ref{fig:bofx}, it appears that adequate foreground cleaning will remove the Fourier modes where primordial non-gaussianity
has its largest impact $k \lesssim 0.02 h$ Mpc$^{-1}$. This rough estimate illustrates the challenge of detecting or constraining
primordial non-gaussianity with redshifted 21 cm measurements, but more detailed calculations are required for a quantitative forecast.
In particular, there may still be a `window' of scales where the distinctive scale-dependent bias signature can be separated from
foreground contamination.

\subsection{Fisher Forecasts}
\label{subsec:fisher}

In order to quantify the prospects for constraining $f_{\rm NL}$ in the
presence of foreground contamination in more detail, we use 
the Fisher matrix formalism. 

\subsubsection{No foregrounds}

We first
consider the constraints that can be achieved in the absence of
foregrounds; the effects of foreground subtraction will be dealt with
below.  We calculate constraints in the two-dimensional parameter
space $\vec{p} = \left(f_{\rm{NL}}, b_x^G \right)$, where $b_x^G$ is the
gaussian bias. In general, $\avg{x_i}$ -- and additional scale-independent 
clustering enhancement terms -- might be included as additional parameters. 
For simplicity, we instead fix $\avg{x_i}$ and stick to this two parameter
model here, since we expect
variations in $b_x^G$ to be more important.
The Fisher matrix is given by
\begin{eqnarray}
\label{eq:fisher}
\mathcal{F}_{\alpha \beta} &=& \frac{1}{2} \mathrm{Tr} \left[C_{,\alpha} C^{-1} C_{,\beta} C^{-1} \right],
\end{eqnarray}
where $C$ is the covariance matrix of the observed data, and $\alpha$,
$\beta$ index the components of $\vec{p}$.\footnote{The fact that the
  mean of our observable vector is zero allows us to write the Fisher
  matrix in this way.} The parameter covariance matrix, $C^p_{\alpha
  \beta} = \avg{p_{\alpha} p_{\beta}}$, (where the average is
over many cosmological realizations) can then be obtained from
$\mathcal{F}$, assuming gaussian statistics, by
\begin{eqnarray}
\left( C^p \right)^{-1} = \mathcal{F},
\end{eqnarray}
where we have used the superscript $p$ to distinguish between the
parameter covariance matrix and the covariance matrix of the observed
data.  Constraints on $f_{\rm{NL}}$ and $b_x^G$ can then be extracted
from $C^p$ in the usual way.

To compute the covariance matrix of the observed data, $C$, we define
a data vector, $\vec{\delta}$, whose components $\delta_i$ represent
the observed 21 cm brightness temperature fluctuations in the voxels
of the experiment.  The data covariance matrix is related to the
observable vector by $C_{ij} = \avg{\delta_i \delta_j}$,
where again the average is over many cosmological realizations.  The
covariance matrix $C_{ij}$ can be written as a sum of a part due to
the signal, $C_S$ and a part due to noise, $C_N$.  

To simplify the calculation of the signal covariance matrix, we ignore
any anisotropy of the power spectrum (due, for instance, to redshift
space distortions), which should be a good approximation during most
of reionization.  We divide the total survey volume into sub-surveys
with line of sight depth corresponding to a bandwidth of 30 MHz (as we
will discuss in a moment, foreground subtraction will be performed
separately in each sub-survey). For simplicity we neglect evolution
across the line of sight depth of each sub-survey. Although this is an imperfect approximation, it
should be adequate for our purposes since
foreground cleaning removes the long wavelength modes along the line
of sight where evolution should be most important.   
Given these
assumptions, the signal covariance matrix in a single sub-survey can
be written as
\begin{eqnarray}
\label{eq:cov}
\left[ C_S \right]_{ij} &=& T_0(z_i)T_0(z_j) \left(\frac{D(z)}{D(0)}\right)^2 \nonumber \\
&& \times \int \frac{d^3k}{(2\pi)^3} \left[ P_{\delta_\rho,\delta_\rho}(k)
  \left(b_{\rm 21}(k,z) \right)^2 \right] \tilde{\psi}_{i}(\vec{k}) \tilde{\psi}^*_{j}(\vec{k}), \nonumber \\
\end{eqnarray}
where we have defined 
\begin{eqnarray}
\label{eq:b21}
b_{\rm 21}(k,z) = -b_x^G(z) - \frac{3(b_x^G(z)-1)f_{\mathrm{NL}}\Omega_{\rm m} H_0^2 \delta_B}{c^2 D(z)k^2T(k)} + (1-\avg{x_i}). \nonumber \\
\end{eqnarray}
In this equation, $b_{\rm 21}(k,z)$ is the bias factor that converts fluctuations in the
underlying density field to fluctuations in a 21-cm map. The form here follows from the term in the square brackets of
Equation \ref{eq:pk21}. The 21 cm bias is negative because large scale overdense regions
are preferentially ionized during reionization -- and consequently dim in 21 cm -- while large scale underdensities remain neutral and are hence bright in 21 cm. 
The $T_0(z)$ factors are dimensionful functions that
perform the unit conversions into 21-cm brightness temperature as in Equation \ref{eq:t21}. $D(z)$
is the linear growth function, $P_{\delta_\rho, \delta_\rho}$ is the
linear matter power spectrum at present day, and $\psi_i(k)$ are the Fourier
transforms of the voxel window functions.  Here $\delta_B$ is related
to the reionization barrier. The precise value of $b_x^G(z)$ and
$\delta_B$ vary with the stage of reionization (especially with
$\avg{x_i}$) and the reionization model.  For simplicity, we ignore
redshift evolution and use fiducial values of $b_x^G(z) = 1.5$,
$\avg{x_i} = 0.3$ and $\delta_B = 1.0$ which are in good agreement
with both the analytic and numerical results presented above.

We further assume that our voxels are spherical top hats of
radius $\Delta r$. In detail, 21 cm surveys will have higher frequency resolution than angular resolution, but primordial non-gaussianity impacts only large
scale fluctuations, and so the higher frequency resolution will not help here. This justifies our use of spherical voxels.
This leads to
\begin{eqnarray}
\tilde{\psi}_i (\vec{k}) &=& \frac{1}{(4/3)\pi (\Delta r)^3} \int_{|\vec{x}| < \Delta r} d^3 x e^{-i\vec{k} \cdot (\vec{x} - \vec{x}_i)} \\
&=& 3e^{-i \vec{k} \cdot \vec{x}_i} \left( \frac{j_1(k_r \Delta r)}{k_r \Delta r} \right),
\end{eqnarray}
where voxel $i$ is located at comoving coordinate $\vec{x}_i$, $k_r$ is
the radial component of $\vec{k}$ and $j_1$ is the spherical Bessel
function of the first kind.  Substituting into Eq. \ref{eq:cov} we
find
\begin{align}\label{eq:csubs}
& \left[ C_S \right]_{ij} = 9 T_0(z_i)T_0(z_j) \left(\frac{D(z)}{D(0)}\right)^2 
\int d \ln k \nonumber \\
& \left[ \frac{k^3}{2\pi^2} P_{\delta_\rho,\delta_\rho}(k) \left(b_{\rm 21}(k,z)\right)^2\right]
 j_0\left(k | \vec{x}_i - \vec{x}_j | \right) \left( \frac{j_1(k\Delta r)}{k\Delta r} \right)^2.
\end{align}

In principle, we need to include contributions from noise, but we
instead consider only $C_s$ here, i.e., we work in the cosmic variance limit. Presently, our
main aim is to explore the {\em fundamental limit} imposed by foreground
cleaning, and so it is appropriate to ignore instrumental noise here.
In this
limit, the normalization factor $T_0(z)$ drops out, and we then have all
the ingredients necessary to compute $\mathcal{F}$ in the absence of
foregrounds.  Below we consider the effects of foreground subtraction
on the Fisher matrix.

\subsubsection{Effects of Foreground Subtraction}

One way to incorporate the effects of foreground subtraction into the
Fisher formalism is to add a large amount of noise to those modes that
are affected by foreground removal (i.e. that are most contaminated by
foregrounds). See \cite{McQuinn:2005hk} and \cite{Liu:2011hh} for related
approaches. Since foregrounds are expected to be smooth along the
line of sight (in the frequency direction) the modes that we consider
here are low-order polynomials along the line of sight and each mode
is non-zero in only a single angular pixel.  In effect, this means
that we are assuming the foregrounds are uncorrelated between
different angular pixels (a conservative assumption).  We will define
$m^a(\vec{\theta}_i, z_j)$ to be the value of the $a$th mode in the
$i$th angular pixel and the $j$th redshift bin.  So, for instance, the
constant mode in the angular pixel labeled by $\alpha$ is
\begin{eqnarray}
m^{constant}(\vec{\theta}_i , z_j) = 
\begin{cases}
1, & \text{for all $j$ if $i = \alpha$}, \\
0, & \text{for all $j$ if $i \neq \alpha$}.
\end{cases}
\end{eqnarray}
Higher order modes correspond to higher order polynomials in $z_j$.

To incorporate mode subtraction into the Fisher formalism, we define a
constraint matrix
\begin{eqnarray}
C_{con} \left(\vec{\theta}_i, z_j, \vec{\theta}_{i'}, z_{j'} \right) = \kappa \sum_{a = 1}^{N_{\rm{modes}}} m^a(\vec{\theta}_i, z_j) m^a(\vec{\theta}_{i'}, z_{j'}), \nonumber \\
\end{eqnarray}
where $m^a_i$ is the value of the $a$th mode in the $i$th voxel and
$\kappa$ is some large number.  The data covariance matrix is then
adjusted by the constraint matrix:
\begin{eqnarray}
C = C_S  + C_{con},
\end{eqnarray}
and the computation of the Fisher matrix proceeds as before.

Ref.\ \cite{Bowman:2008mk} has suggested that foregrounds can be fit with a
cubic polynomial over a bandwidth of $\sim 30$ MHz.  We therefore
consider the effects of foreground subtraction separately in each
subsurvey. We explore the effects of foreground subtraction at this
level, and also vary the number of foreground modes (using higher
order polynomials) to better understand the robustness of constraints on
$f_{\rm{NL}}$ to foreground subtraction.

\subsubsection{Fisher Results}

In order to quantify the constraints on $f_{\rm NL}$ that can be
obtained in the future using the technique presented here, we consider
the case of a full-sky survey that covers a frequency range from 120
Mhz to 210 Mhz. This survey covers redshifts between $z=5.8$ and
$z=10.8$, centered on $\avg{z} = 8.3$.\footnote{The Universe is likely
  fully ionized at the low redshift end of the range considered here,
  but we don't expect our results to depend sensitively on the precise
  redshift range considered.}  Both the survey area and the large
bandwidth of our hypothetical survey are optimistic, but this is
appropriate for exploring the ultimate limits on $f_{\rm NL}$
constraints from the reionization-era 21 cm signal.  As discussed
above, the survey volume is divided along the line of sight into three
subvolumes, each with depth 30 Mhz.  We divide the survey region into
voxels measuring $\frac{1}{3}^{\circ} \times \frac{1}{3}^{\circ}$
across the sky and 3 Mhz in the frequency direction.  The resulting
voxels are roughly cubical with side length (diameter) 50 Mpc so that
our spherical voxel assumption is not a bad approximation.  The
effects of using finer voxelizations are explored below.

As our fiducial survey has several million voxels, the resulting
covariance matrix is very large.  Rather than attempt to invert this
large matrix, we compute the Fisher matrix for a
$5^{\circ}\times5^{\circ}$ region (with equivalently sized voxels),
and scale the resulting Fisher matrix to account for greater sky
coverage.  Computing the Fisher matrix in this way assumes that no
information is contributed by pairs of voxels separated by more than
$5^{\circ}$. This is a conservative assumption and we expect
it to be a reasonable approximation. We assume
that foreground removal is performed separately for each line of sight
and for each of the three subvolumes of the full survey.

\begin{figure}
\begin{centering}
\includegraphics[width=1.0\columnwidth]{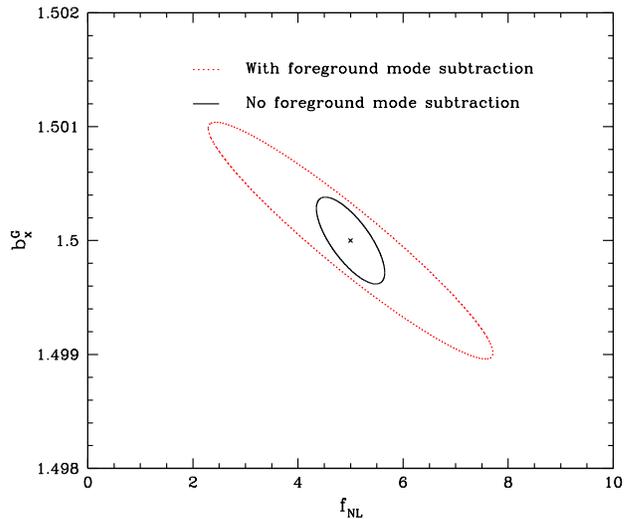}
\caption{Fisher projections for the error on $f_{\mathrm{NL}}$ and
  $b_x^G$ with and without foreground mode subtraction in the two
  dimensional parameter space defined by $f_{\rm{NL}}$ and $b_G$.
  The cross shows our fiducial parameter values. The survey considered
here is a futuristic, full-sky, cosmic-variance limited 21 cm survey
covering 120-210 Mhz.}
\label{fig:fisher_2d}
\end{centering}
\end{figure}

Figure \ref{fig:fisher_2d} shows the projected constraints for the
survey described above in the two dimensional parameter space defined
by $\vec{p} = \left(f_{\rm{NL}}, b_x^G \right)$.  Fiducial values of
$f_{\mathrm{NL}}$ and $b_x^G$ are marked with a cross-hair; they are
$(f_{\mathrm{NL}},b_x^G) = (5,1.5)$.  We show both the constraints
obtained in the absence of foregrounds (red, dotted curves) and those obtained
when foreground subtraction is implemented in the manner described
above (black, solid curve). Although the foregrounds degrade the detection,
it seems that significant constraints are still reasonable even after
foregrounds are subtracted. Specifically, the 1-$\sigma$ constraints
on $f_{\rm NL}$ after marginalizing over $b_x^G$ are 0.43 and 1.8 before and after removing foreground
contaminated modes, respectively.

To get a better sense of the range of scales that contribute most to
the predicted constraint on $f_{\rm{NL}}$ in the presence of
foregrounds, we now consider the effects of varying the number of
voxels used in the experiment and also the number of foreground modes
that are subtracted.  Decreasing the number of voxels effectively
decreases $k_{max}$, the maximum wavenumber used in the constraint,
while increasing the number of foreground modes effectively increases
$k_{min}$.  Here, rather than consider an experiment across the full
bandwidth, we consider one of the sub-surveys, ranging from 120 Mhz to
150 Mhz.

\begin{figure}
\begin{centering}
\includegraphics[width=1.0\columnwidth]{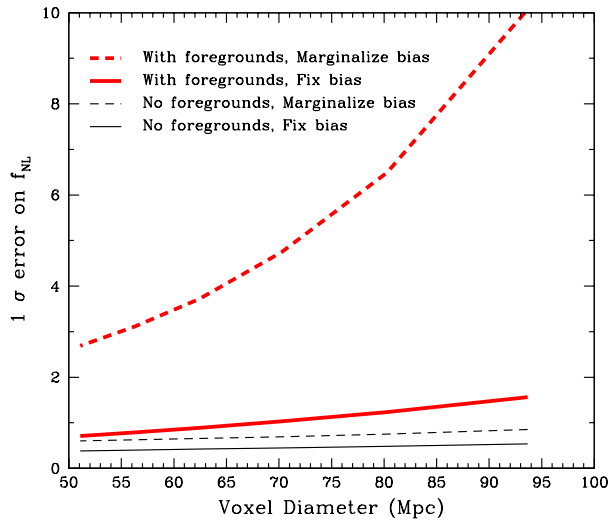}
\caption{Fisher projections for the error on $f_{\mathrm{NL}}$ and
  $b_x^G$ with and without foreground mode subtraction as a function of
  the voxel radius of the experiment. The projections here are for a
  survey with 30 Mhz bandwidth rather than the 90 Mhz bandwidth
  assumed in Fig.~\ref{fig:fisher_2d}.}
\label{fig:fisher_pix_converge}
\end{centering}
\end{figure}

Figure \ref{fig:fisher_pix_converge} shows the projected errors on
$f_{\mathrm{NL}}$ as a function of the voxel size of the survey.  In
generating this figure we have assumed the same fiducial values of
$f_{\rm{NL}}$ and $b_x^G$ as above: $(f_{\mathrm{NL}},b_x^G) =
(5,1.5)$.  It is clear that the $1\sigma$ error on $f_{\rm{NL}}$
declines rapidly with decreasing voxel size until the voxel diameter is
roughly 50 Mpc; little information is gained by using smaller voxels.
This means that most of the information on $f_{\rm NL}$ is coming from
large scales, as anticipated.

\begin{figure}
\begin{centering}
\includegraphics[width=1.0\columnwidth]{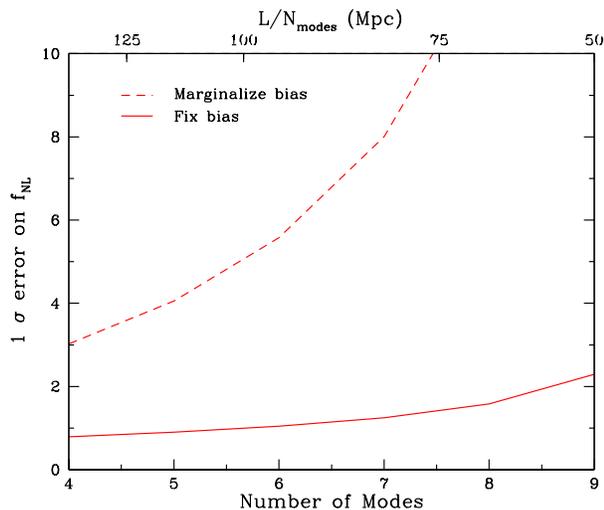}
\caption{Fisher projections for the $1\sigma$ error on
  $f_{\mathrm{NL}}$ as a function of the number of foreground modes
  that are subtracted (over a bandwidth of 30 MHz, with 50 Mpc
  diameter voxels). Each foreground mode is a polynomial so that
  e.g. subtracting 6 modes corresponds to fitting a 5th order
  polynomial across the bandwidth of the experiment. Based on previous
  studies $N=4$ is expected to be sufficient to clean foregrounds.}
\label{fig:fisher_mode_converge}
\end{centering}
\end{figure}

We can get a handle on the maximum scale that 
contributes to the $f_{\rm{NL}}$ constraint after foreground cleaning
by varying the number of foregrounds modes
that are subtracted.  Figure \ref{fig:fisher_mode_converge} shows the
projected errors on $f_{\mathrm{NL}}$ as a function of foreground
modes that we subtract across the 30 Mhz bandwidth.  The upper x-axis
shows the corresponding maximum scale that can be constrained by the
data, given by $L/N_{\rm{modes}}$, where $N_{\rm{modes}}$ is the
number of foreground modes subtracted, and $L$ is the dimension of the
survey.  Based on previous studies (e.g. \cite{Bowman:2008mk}), we expect
$N=4$ modes to be sufficient to remove foregrounds over the bandwidth
considered. Provided
this is indeed sufficient, there should be a narrow range of scales that
are impacted significantly by $f_{\rm NL}$, yet survive foreground cleaning.
In particular, Figures \ref{fig:fisher_pix_converge} and \ref{fig:fisher_mode_converge},
demonstrate that the constraint on $f_{\rm NL}$ comes mostly from the narrow range of scales
between 50 Mpc to 125 Mpc, corresponding to roughly $k \sim 0.01-0.03~h \rm{Mpc}^{-1}$.
The ability to tightly constrain $f_{\rm NL}$, in spite of this limited dynamic range in scale, is
driven by the large volume of our futuristic survey. This survey samples many modes on the
scales of interest and thereby provides small statistical errors on the power spectrum. It is encouraging
that this may, in principle, allow constraints that are competitive with -- or even slightly better than  -- 
Planck (as seen in Figure
\ref{fig:fisher_2d}). The 1-$\sigma$ error in Figure \ref{fig:fisher_2d} is $1.8$, which compares favorably to the existing 1-$\sigma$ error from Planck of $5.8$ \cite{Ade:2013ydc}. However, because of foreground cleaning, the 
$1/(k^2 T(k))$ signature will not be observed over a large range in scales and will imprint only
a slight excess power in the largest scale modes. In effort to ensure that this signature 
is robust to foreground
cleaning, one might test sensitivity to the number of foreground modes that are projected-out, along
the lines of Figure \ref{fig:fisher_mode_converge}.

\section{Discussion and Conclusions}\label{sec:conclusions}

In this work we have investigated the effect of primordial
non-gaussianity on the bias of ionized regions during reionization. We
have extended the analytic model of Furlanetto
et.\ al.\ \cite{Furlanetto:2004nh} to the case of non-gaussian initial
conditions and demonstrated that ionized regions will exhibit a scale
dependent bias. Semi-numeric simulations confirm these results.

We have investigated the constraints that measurements of the power
spectrum of fluctuations in the 21 cm radiation from the Epoch of
Reionization may place on the gaussianity of the initial
conditions. We find that futuristic redshifted 21 cm surveys might allow
constraints that are competitive with Planck, in spite of foreground cleaning.
This is still going to be a challenging endeavor, since interesting values 
of $f_{\rm NL}$ lead to only small changes in the power spectrum
for the modes that survive foreground cleaning. Future galaxy surveys will also face severe
systematic challenges in order to robustly measure the large scale galaxy power spectrum and constrain $f_{\rm NL}$ (e.g. \cite{Ross:2012sx}). The
future prospects also need to be considered in the context of the recent 
Planck constraints, which almost close the door on
the prospects of detecting primordial non-gaussianity, and using this to
probe the physics of inflation.
Nevertheless, upcoming galaxy surveys and redshifted 21 cm measurements
will survey tremendous volumes, and can measure the large scale power spectrum with high statistical
precision. These upcoming measurements are interesting in their own right, and will proceed
regardless of searches for primordial non-gaussianity. They still deserve careful scrutiny since
they should be precise enough to reveal subtle signatures from small levels of
primordial non-gaussianity. Galaxy and 21 cm surveys suffer from different
systematic concerns, and their combination may still provide a powerful
test of primordial non-gaussianity in the post-Planck era.

\acknowledgements

This work was supported in part by National Science Foundation under
Grant AST-090872, the Kavli Institute for Cosmological Physics at the
University of Chicago through grants NSF PHY-0114422 and NSF
PHY-0551142 and an endowment from the Kavli Foundation and its founder
Fred Kavli. AL was supported in part by the NSF
through grant AST-1109156. SD is supported by the U.S.  Department of Energy,
including grant DE-FG02-95ER40896.   PA thanks the Kavli Institute for Theoretical Physics for hospitality and support through National Science Foundation Grant No. NSF PHY11-25915 as this work was nearing completion.

\bibliographystyle{hieeetr}

\bibliography{references}

\end{document}